# Characterization of Intact Eukaryotic Cells with Subcellular Spatial Resolution by Photothermal-Induced Resonance Infrared Spectroscopy and Imaging.

Luca Quaroni [1*]

[1] Faculty of Chemistry, Jagiellonian University, ul. Gronostajowa 2, 30-387, Kraków, Poland; luca.quaroni@uj.edu.pl

[2] Institute of Nuclear Physics, Polish Academy of Sciences, ul. Radzikowskiego 152 31-342 Kraków, Poland

* Correspondence: luca.quaroni@uj.edu.pl

**Abstract:** Photothermal-Induced Resonance (PTIR) spectroscopy and imaging with infrared light has seen increasing application in molecular spectroscopy of biological samples. The appeal of the technique lies in its capability to provide information about IR light absorption at a spatial resolution better than allowed by light diffraction, typically below 100 nm. In the present work we test the capability of the technique to perform measurements with subcellular resolution on intact eukaryotic cells, without drying or fixing. We demonstrate the possibility to obtain PTIR images and spectra from the nucleus and multiple organelles with high resolution. We obtain particularly strong signal from bands typically assigned to acyl lipids and proteins. We also show that while a stronger signal is obtained from some subcellular structures, other large subcellular components provide a weaker or undetectable PTIR response. The mechanism that underlies such variability in response is presently unclear. We propose and discuss different possibilities, addressing thermomechanical, geometrical and electrical properties of the sample and the presence of cellular water, from which the difference in response may arise.

## 1. Introduction

Infrared absorption spectroscopy in the mid infrared spectral range (approximately 2.5 – 25 μm or 4000 - 400 cm-1) is a valuable technique for investigating molecular properties in soft matter research, including samples of biochemical and biological interest. Over the last few decades it has been used extensively for investigating composition, structure and reactivity of a wide range of samples, ranging in complexity from purified biomolecules to tissue sections. The introduction of microscopes for IR spectroscopy and imaging has extended applications to the spatially resolved analysis of sample heterogeneity on the micrometric scale. IR microscopes have been successfully used with tissue sections and biopsies, and in the spectroscopic study of single eukaryotic cells. However, the use of far-field optics has constrained the spatial resolution of the measurement to the values imposed by diffraction, of the order of the wavelength in the mid-IR spectral region. The size of most prokaryotic cells is below the shortest wavelength, while most eukaryotic cells range in size between 10 and 100 μm. As a result, most IR microscopy studies have been restricted to performing spectroscopy of single whole eukaryotic cells. Studies with subcellular resolution have been few and limited to larger cells, often requiring the use of synchrotron radiation to improve signal-to-noise at diffraction limited resolution. In most cases such studies allow selectively probing the nucleus and the larger vacuoles, but most subcellular structures remained unresolved.

The introduction of techniques for nanoscale IR spectroscopy has promised to open the way to measurements with resolution better than allowed by optical diffraction. Several designs have been introduced that allow a resolution better than λ and often better than 1/10 λ in the mid IR spectral region, including transmission-mode Scanning Near Field Optical Microscopy (SNOM),[1,2] scattering-mode SNOM,[3] photothermal induced resonance (PTIR),[4] Photothermal Microspectroscopy (PTMS),[5] scanning thermal infrared microscopy (STIRM),[6] photoinduced force microscopy (PiFM).[7] Among these techniques, PTIR detects absorption of IR light by relying on the deflection of an AFM probe following photothermal expansion of the sample in the contact location. PTIR was originally introduced by Dazzi et al. [4] and was based on the use of CO2 and Free Electron Laser (FEL) sources for excitation. The later introduction of a more extensive park of benchtop light sources, including Optical Parametric Oscillator (OPO) and Quantum Cascade Lasers (QCL), allowed extending the accessible wavenumber region and introducing novel experimental configurations that made the technique more accessible and applicable to a wider sample range.[8] As a result, PTIR has seen increased application in material characterization, particularly concerning soft matter samples. These developments have been recently reviewed. It is generally accepted that PTIR can be successfully used for the study of single cells with a resolution better than 100 nm.[8] Samples that have been investigated by PTIR to date include both prokaryotic cells [9–12] and eukaryotic cells.[13–20] Among the latter, one example includes the study of living yeast.[16] To our knowledge, the latter is the only reported example of PTIR measurements on a single living cell in an aqueous environment to date. The reason for the limited amount of work done on the subject can be ascribed to the difficulty of performing PTIR measurements in an aqueous solution. Complications arise from a combination of factors that include the high absorption of water in the mid-IR region, the need to retain the alignment of a light beam that crosses the air-water interface, and the general damping of cantilever resonances in a liquid medium. [21] Optical and mechanical limitations make even the measurement of simple molecules challenging. The use of an aqueous environment is often considered a necessary condition for the investigation of living cells, because it reproduces the typical conditions used in cell culture. However, while water retention in a eukaryotic cell maintains cellular viability, the presence of a bulk aqueous environment at the exterior of the cell is not strictly necessary. Living cells in vertebrate tissue exist in three-dimensional ensembles that have no direct contact with a bulk aqueous phase. Only few of them exist in aqueous suspension, notably blood cells, while more exist in tissue layers in contact with an aqueous phase, such as epithelial cells in blood vessels. It is therefore of interest to explore the feasibility of PTIR measurements on eukaryotic cells in the absence of a bulk aqueous phase but without drying of the cells. We want to assess the capability of the sample to retain its structure despite the mechanical stress associated to AFM scanning in contact mode and the thermal stress associated to PTIR measurements. In the present work we use buccal cheek cells because of their accessibility and because of their stability when exposed to air. The cells can be collected together with their coating of salivary biopolymers, which are involved in the retention of humidity in the mouth cavity and contribute to the stabilization of the sample in the atmosphere. We perform both spectromicroscopy and imaging experiments on the cells and demonstrate the capability to resolve subcellular structures less than a micrometer in size.

## 2. Results

We used PTIR in contact mode to obtain images and single spectra of freshly collected buccal epithelial cells. Cellular topography was characterized by recording AFM images in contact mode. The cells responded well to contact scans, without major structural changes and were mechanically stable over multiple AFM imaging scans. When measured with high resolution, either in AFM and/or in PTIR images, (under irradiation with IR light), the cells displayed only slow changes in morphology over time, within the range of sample drift. This morphological evolution appears to be the result of slow drying and mechanical settling of the cell, rather than of mechanical damage caused by scanning, as already remarked. We then recorded multiple parallel AFM and PTIR images of the sample ranging from a full cell to subcellular regions of decreasing size. Figure 1 shows an AFM image and corresponding PTIR images of a single whole cell and of a subcellular portion. The wavelengths for photothermal excitation were chosen to match the known absorption bands of cellular samples between 1700 cm$^{-1}$ and 1000 cm$^{-1}$.

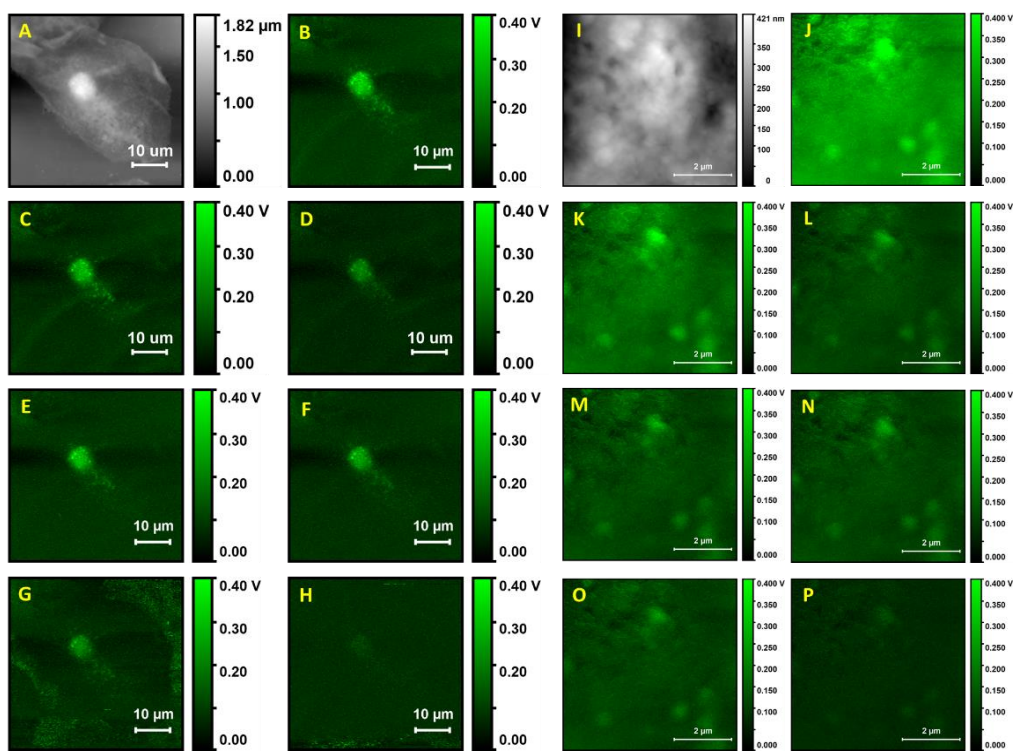

**Figure 1.** AFM and PTIR images of a single hydrated buccal cell. A) AFM height map of whole cell. B) PTIR map of cell at 1678 cm$^{-1}$, C) PTIR map at 1650 cm$^{-1}$, D) PTIR map at 1545 cm$^{-1}$, E) PTIR map at 1440 cm$^{-1}$, F) PTIR map at 1380 cm$^{-1}$, G) PTIR map at 1275 cm$^{-1}$, H) PTIR map at 1090 cm$^{-1}$. I) AFM height map of cell detail. J) PTIR map of cell detail at 1650 cm$^{-1}$, K) PTIR map at 1545 cm$^{-1}$, L) PTIR map at 1440 cm$^{-1}$, M) PTIR map at 1380 cm$^{-1}$, N) PTIR map at 1275 cm$^{-1}$, O) PTIR map at 1090 cm$^{-1}$. PTIR Maps B, C, D and J, K, L were measured at 8% of maximum power, other maps at 17% of maximum power.

The overview of a cell topography by contact mode AFM is shown in Figure 1a. The cell appears to be flattened on the surface of the support. The nucleus can be easily identified because it gives rise to an ellipsoidal protrusion 8-10 μm in size. Similarly to the case of adherent cells, the nuclear location corresponds to the position of maximum thickness for the cell, about 1.8 μm in our case. From this position the cell body decreases progressively in thickness towards the edges. Much of the cell body appears to be between 0.6 and 1.2 μm in thickness. Overall the cell has a granular appearance and inspection of a portion at higher resolution (Figure 1J) shows a rough and puckered surface, with rugosity of the order of 100 - 200 nm. Except for the nucleus, no other subcellular structures can be clearly discerned from the topography image.

PTIR images of the same cell were recorded while exciting the sample at different wavelengths, corresponding to different absorption peaks. The nucleus is clearly observed with excellent contrast in images collected at most wavelengths, except at 1090 cm$^{-1}$, which provides a weak image. In addition to the nucleus, a strong signal is observed from particles with a spheroidal or occasionally oblong structure, about 0.5 to 1.5 μm in size, which can also be observed at all excitation wavelengths. Because of size, abundance and cellular distribution, the structures are likely identified as organelles, probably peroxisomes or lysosomes, although we cannot rule out mitochondria or lipid droplets. Except for these structures, very weak signals are obtained from the remainder of the cell. The signal decreases rapidly away from the nucleus, at all exciting wavelengths, and is close to zero mV throughout much of the cell volume.

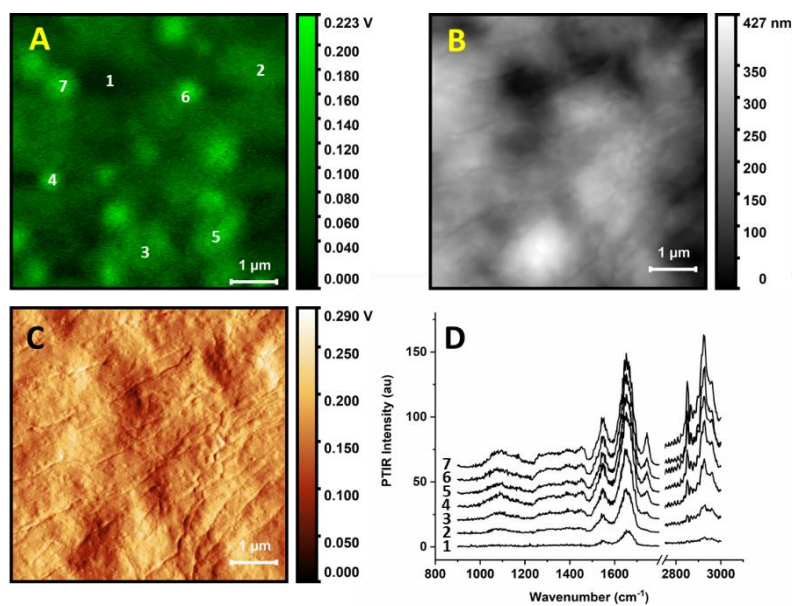

**Figure 2.** AFM maps, PTIR maps and single spectra of subcellular locations. A) PTIR image of portion of a buccal cell collected with 1670 cm$^{-1}$ excitation. B) AFM height image of the same cellular portion as in A. C) Deflection image of the same cellular portion as in A. D) PTIR spectra recorded in different locations of the sample region, numbered according to panel A. Power was set at 20% of the maximum.

Figure 2 shows spectra collected at single locations of a cell (this is a different cell than the one used in Figure 1). AFM and PTIR images were used to locate an area of interest, in this case a portion of a cell away from the nucleus. The region allows us to investigate multiple organelles in a single frame, as well as the space between them. The PTIR map at 1670 cm$^{-1}$ does not show any obvious correspondence with the height and deflection maps, showing that contributions from topography, even if present, are not dominating the PTIR signal. The probe was located at the selected position and spectra were recorded using the second resonance of the cantilever (Figure 2D).

Repeated measurements of the sample over several hours (not shown) show little change in overall structure of PTIR images, with the only change being a slow decrease of signal intensity, although the cell does not appear to dry out.

All PTIR spectra show a similar pattern of bands. The strongest bands are observed at 1650 cm$^{-1}$ and 1545 cm$^{-1}$, as is common in the IR absorption spectra of single cells. This region is typically dominated by a strong contribution from the Amide I and Amide II band doublet of polypeptides and amides, plus water, although contributions from amines, carboxylate and other carbonyl groups are also present. Weaker absorption bands are observed at lower wavenumbers, down to 1050-1100 cm$^{-1}$. Most spectra display similar band patterns and differ mostly in overall intensity. The main clear variable between spectra is a set of peaks at 1740 cm$^{-1}$, 2850 cm$^{-1}$ and 2925 cm$^{-1}$. The bands are present in all spectra measured on organelle particles but are not seen when measuring away from them. In cellular IR spectra these bands typically arise from the long chain acyl groups of fatty acids and lipids, including phospholipids and triglycerides, although overlap with bands from other molecular components is also possible. Remarkably, other bands arising from acyl lipids, such as the CH$_2$ bending mode around 1460 cm$^{-1}$ and headgroup vibrations around 1050 cm$^{-1}$ - 1090 cm$^{-1}$ are not obvious and may be present only as weak or unresolved components of the broad featureless bands observed below 1500 cm$^{-1}$. Stronger spectra arise from the brighter spots observed in the PTIR map. In contrast, spectra measured away from the organelles are much weaker and allow discerning only the doublet of peaks in the Amide region (Also see supplementary information for more examples of spectra). The intensity of the spectra does not seem to be clearly related to the thickness of the sample: positions 4, 5, 6 and 7 give the strongest spectra despite being as thick as positions 2 and 3. Figure 3 highlights this observation by showing matched profiles along selected locations of a height map and the corresponding PTIR map from Figure 1. The stronger spikes on top of the two PTIR profiles are signals from organelles.

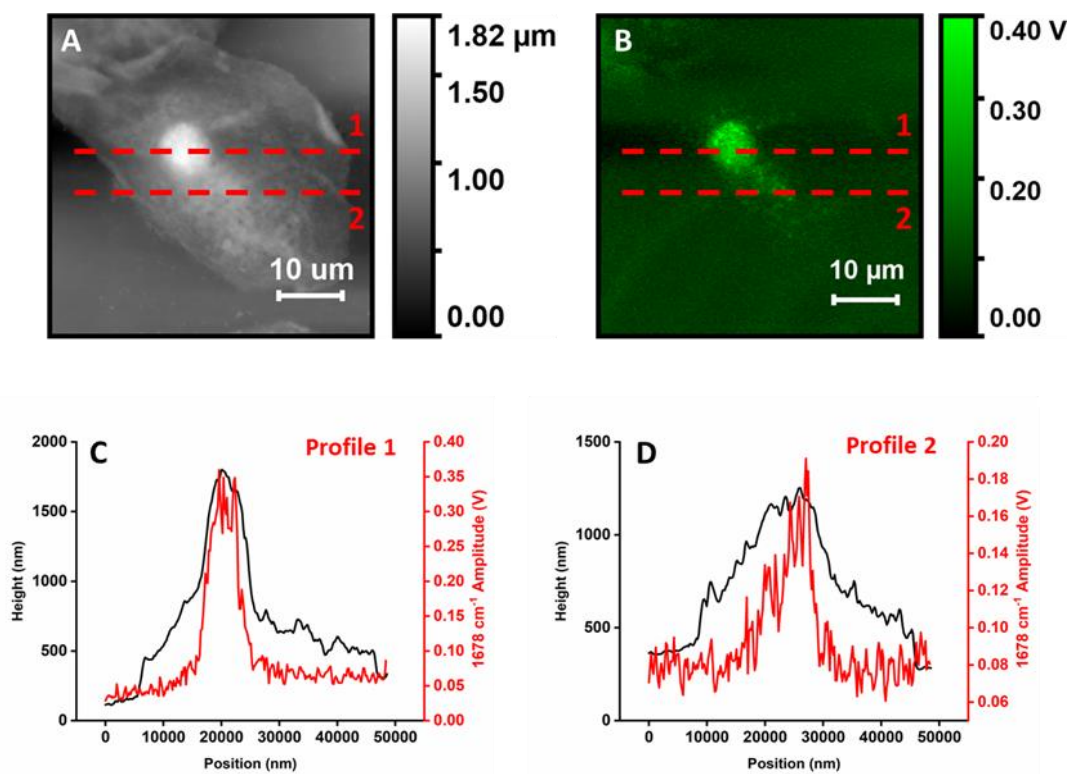

**Figure 3.** Comparison of Height and PTIR Amplitude profiles with excitation at 1678 cm-1 for a single cell. A – Height map. B – PTIR map with 1678 cm-1 excitation, collected in synchrony with A. C – Profiles along line 1 in A and B. D - Profiles along line 2 in A and B.

## 3. Discussion

Buccal cheek epithelial cells are components of the mouth epithelium that are continually shed into the saliva. The cells are harvested either from the saliva or directly by scraping the epithelium in a variety of states, ranging from intact to micronucleated, binucleated, pyknotic or karyolitic. The cells can be successfully transferred to culture conditions, indicating that at least some of them are generally viable. [22,23] When tested by Trypan Blue, a large number of cells shows dye exclusion, indicating that the membrane is generally intact. However, dye exclusion tests on these cells are also known to give ambiguous results, with some live cells showing dye uptake [23] and we cannot confirm whether the individual cells investigated in this work are actually alive. We avoid cells with irregular appearance and fragmented edges and select cells with a visible nucleus which is at least 8-10 μm in size, to exclude karyolitic, necrotic and binucleated cells. Even when exposed to the atmosphere in the course of the measurement the cells dry out slowly and retain their morphology for a few hours and over repeated AFM and PTIR scans. The stability of this cellular system in air is remarkable and permits collecting multiple AFM/PTIR images of the same cell at different excitation wavelengths without loss of structure. A measurable PTIR signal is obtained even at relatively modest excitation power (~0.1 - 1.0 mW). Together with the stability of the sample, the response permits collecting multiple PTIR images at different wavelengths while retaining comparable structure.

Supporting Figure 1 compares the results for two separate cells of different thickness and structure. Comparison of the PTIR signal at the nuclei shows that the intensity is proportional to the height of the nucleus. For bulk compounds, the PTIR signal is expected to increase with the thickness of the sample, as shown experimentally in the case of polymethylmethacrylate (PMMA), for which the signal shows a linear increase with sample thickness, up to about 1 µm, followed by a decrease.[24] However, complications are observed for layered materials, where the mechanical properties of the tip-sample contact affect signal intensity.[25] In our experiments, the linearity of the response with thickness is confirmed when looking at nuclei of different thickness (See Supporting Information and Supporting Figure 1). However, the relationship between sample thickness and signal intensity is no longer linear when comparing different subcellular regions. Comparison of PTIR and AFM topography maps of Figure 1 show that the PTIR signal is maximal at the nucleus, corresponding to about 1.8 µm in thickness, but is minimal or absent in other locations of the cell that are as thick as 1 µm. The organelles are about 0.5 to 1 µm thick but provide a stronger signal than cellular regions of comparable or greater thickness. Figure 3 uses line profiles to highlight that signal intensity does not increase linearly with the thickness of the cell but is localized mostly at spherical or spheroidal structures. Apart from nucleus and organelles, the only other visible structures, although with weaker contrast, are pits and protuberances in the cellular surface that are apparent in higher-resolution higher-intensity images (Figure 1 J-M). The latter structures arise from changes in the contact frequency of the tip as it scans through regions of the sample with different mechanical properties (see Supporting Information and Supporting Figure 2). Changes in the Young modulus of microscopic and nanoscopic regions of the surface, as revealed by changes in contact frequency, correspond to changes in PTIR signal that modulate the intensity of spectral bands and of PTIR maps.[26] This is further detailed in the Supporting Information (Supporting Figure 3 and associated paragraph). Overall, changes in the contact frequency introduce fine patterns in PTIR images, often with very high spatial resolution, which do not necessarily correspond to changes in the IR absorption of the sample. While these patterns contain interesting information about the micromechanical properties of the surface of the sample, they will not be addressed further in this work and will be the subject of future investigations.

The chemical composition of a eukaryotic cell is complex and inhomogeneous and varies throughout the location within the cell. Excitation at different wavelengths, corresponding to the absorption maxima of different molecular components, is expected to generate contrast based on the varying chemical composition and molecular properties of subcellular regions. This was indeed observed in our recent work on fixed fibroblast cells, where we could selectively image components of the cytoskeleton, lipid droplets, vesicles, and fragments of organelles by changing excitation wavelength.[18] Surprisingly, we cannot observe the same variety of structures in the present work. Overall it appears that the largest PTIR signals are produced by spheroidal organelles and are dominated by the bands typically assigned to acyl lipids and polypeptides, although other contributions are possible. Weaker signals come from the nucleus and seem to arise mostly from bands in the amide absorption region. Little contribution is seen on the nucleus from bands of acyl lipids, in agreement with what we know about its composition of the nucleus and with far-field FTIR studies of nucleated eukaryotic cells. Other components expected to give significant IR absorption, such as the larger cytoskeletal structures, are barely visible or not observed at all.

The relative lack of content in PTIR images is confirmed by spectromicroscopy measurements at specific locations. Strong spectra are observed when measuring on top of the organelles, with the band patterns expected from particles that are rich in both proteins and phospholipids. In contrast, much weaker spectra are measured elsewhere. In contrast to PTIR imaging measurements, PTIR spectra recorded on the nucleus, where the only obvious bands are the ones conventionally assigned to polypeptides and the overlapping water absorption, also show weak absorption bands. Some instances of PTIR measurements on fixed and dried cells [18] and on proteins [27] have provided extremely skewed Amide I to Amide II ratios, where the Amide II band is nearly undetected and/or where the Amide I band is unusually sharp. None of these effects, so far still unexplained, are seen in the spectra reported in this study.

Part of the difference between the PTIR spectra and maps of fixed and intact cells is likely due to the treatment that accompanied fixation. Fixed samples from our previous experiments had also been treated with Triton X-100 to disaggregate the cellular membrane. The treatment also appeared to degrade most organelles. It is also likely that some of the larger lipid and protein rich particles that were observed in fixed cells (e.g. the lipid coated spheroid with a protein core shown in Figure 4 of reference [18]) are the remains of those of the organelles that survived fixation. Other differences, such as the difficulty in detecting cytoskeletal components, may arise from the physiology of the different cell types. The fixed cells studied in our previous work were fibroblasts, which display a more extensive cytoskeletal network related to their function.

One interesting question is the contribution of water absorption to PTIR spectra, particularly in the 1640-1650 cm-1 region. In far field FTIR measurements of aqueous samples this region typically shows saturation or higher noise levels due to decreased light throughput, hindering measurements of samples thicker than 10 μm. In the present samples, avoiding the use of a bulk aqueous phase reduces this contribution to that of intracellular water (50% to 70% of cell mass, depending on cell type), in cells that are between 1 and 2 μm at the thickest location. The contribution from water to the PTIR spectra of these cells, which falls at 1645 cm-1 in far-field IR absorption spectra of bulk water, is surprisingly small, even though the cells were not dried. Water is expected to contribute to cellular PTIR spectra in two ways. One is absorption of the incident laser beam itself by cellular water, which would lead to a drop in the intensity of the PTIR signal around 1645 cm-1 because of decreased incident power. For thicker samples or for samples in an aqueous medium this would lead to saturation effects and is one of the main complicating factors in performing PTIR in an aqueous environment. However, this is not expected in our case since cells are in air and thinner than 2 μm. The other contribution would come from the photothermal expansion of water itself and would give a positive contribution to the PTIR signal. For an intact hydrated cell this is expected to be comparable to the contribution of other cellular components. This is not observed. The only indication that water may contribute to the PTIR spectrum is that the apparent ratio of bands at 1650 cm-1 and 1545 cm-1 is qualitatively higher than is commonly observed in the FTIR of fixed cells. One possible reason for the absence of a strong water contribution is that absorption at 1645 cm-1 is observed for bulk water, while the structure of water inside a cell is believed to be substantially different from and more ordered than that of the bulk liquid phase.[28] Cytoplasmic water is generally described as a complex web of interconnected water molecules that include the hydration sphere of all biomolecules within the cell and pockets of confined micro and nanoscopic water clusters. Detailed studies of the water bending mode for intracellular water have not been performed. Our previous studies on water turnover in cells in a heavy water medium suggest that it falls in the 1600 cm-1 - 1700 cm-1 region, since it cannot be resolved from the Amide I mode.[29] IR spectroscopy studies of this band for nanostructured water in biological macromolecules[30], nano-IR spectroscopy of confined water[31,32] and spectroscopic[33] and theoretical studies of solvated protons [34] have also shown that absorption of this mode can vary over tens of wavenumbers, depending on the microenvironment and molecular interactions of specific water molecules. The spread can lead to a broad and featureless absorption which may be hard to resolve from other contributions. Specific experiments are necessary to explain water IR absorption or lack thereof, in the cytoplasm of cells.

It is surprising to see such strong and specific response from the membrane components of single organelles under the measurement conditions used in this experiment. Mitochondria are enclosed by two membranes, while lysosomes and peroxisomes are enclosed by one membrane, corresponding to a phospholipid bilayer and associated proteins. Even accounting for the folding of the inner membrane of mitochondria into stacks of cristae, which increases the effective thickness, this is too small an amount of material and is not expected to give an easily detected signal. The measurement of monomolecular layers on gold (roughly equivalent to one leaflet of a membrane) requires the combined contribution of resonant mode PTIR, enhancement by optical coupling between a sharp metallic tip and a gold substrate, and a laser power of the order of 1 kW/cm2, none of which correspond to our experiment. A possible explanation is that the signal is affected by the heterogeneity in chemical composition of the sample and the difference in composition between the content of the organelles and other subcellular structures. Peroxisomes, lysosomes and mitochondria are all involved in lipid and protein metabolism and catabolism and accumulate acyl lipids and proteins at high density. Other metabolites that are turned over by the organelles and can contribute to the observed spectra are small molecules containing amines or protonated carboxylic acid groups.

There are multiple conceivable explanations, not necessarily exclusive. One possibility is that the signal is modulated by the different mechanical properties of the organelles and the nucleus when compared to other structures. The mechanics of tip-sample interaction, as defined by the Young modulus, affect signal intensity [26], with changes in stiffness giving rise to changes in signal intensity.[25] Other thermomechanical parameters of the sample are predicted to affect the PTIR response, such as density, thermal conductivity and heath capacity.[26,35] It is known that some of these quantities are different from organelle to organelle, which allows their separation by centrifugation with a density gradient. Differential distribution of water within the cell is another factor that can have a major impact in the thermal capacity and thermal conductivity of an organelle and of its surrounding environment and can explain the different contrast observed between dry or fixed cells and intact cells which still retain a degree of hydration.

A contribution to the PTIR signal, not considered in existing theoretical treatments, may come from the electrical properties of the sample. Biological membranes act as capacitors, accumulating charges on opposite leaflets in the form of electrolyte distribution, a property which is exploited for energy storage and signal transduction within the cell. In our samples, water loss during the measurement has plausibly affected the distribution and effective concentration of electrolytes and their mobility, but without removing them from the cell. Photothermal expansion of membrane lipids and membrane proteins can affect the capacitance of membranes, which in turn would perturb the electrolyte distribution in the cytoplasm. AFM probes are known to be sensitive to changes in the electrical properties of the sample and AFM technology has been used to measure charge distribution, electrical potential and capacitance with high resolution. It is notable that the tip used in the present experiments, made of silicon with 20 nm gold coating, has similar design as the one used in the original demonstration of Kelvin probe microscopy.[36] However, an explanation based on changes in membrane capacitance does not account for the lack of a clear contribution from membranes of the Golgi and endoplasmic reticulum. In addition, the presence of adsorbed water on the sample surface, as in our samples, degrades contrast in Kelvin probe microscopy [37] and would lead to poorer contrast in intact cells than in fixed and dried cells. Based on the latter arguments, an electrodynamical interpretation of the effect looks unlikely.

Finally, the geometry of the observed structures could also affect signal generation. IR microscopy measurements with far-field optics are affected by distortion of spectral line shapes away from a purely absorptive profile because of optical effects that involve the interplay of the real and imaginary parts of the refractive index of the sample. [38] Most commonly, the distortion results in absorption bands taking on a derivative like profile when measuring objects with a spherical shape. No such distortions are obvious in the spectra reported in this work. This is expected, since IR absorption measurements that rely on the photothermal effect, including PTIR photoacoustic measurements, are generally free of contributions from the real refractive index.[39] However, the present experiments do show a stronger signal arising from structures with a spherical or spheroidal shape. The apparent correlation between shape and signal intensity leaves open the possibility that sample geometry may affect the response via a hitherto unknown mechanism.

With the present data set it is not possible to discriminate between the various hypothesis. It may eventually be possible in future work by appropriate experimental design and by the complementary use of modelling and computations.

**4. Materials and Methods**

PTIR measurements were performed on a nanoIR2 instrument (Anasys, Santa Barbara, CA, USA) working in contact mode using PR-EX-nIR probes. The probes are manufactured out of Au-coated silicon, have a cantilever with force constant 0.007 – 0.4 N/m, nominal tip diameter of 30 nm and eigen frequency equal to 12.79 ± 0.64 kHz in air. The movement of the AFM laser was used to record the PTIR signal as an oscillatory decay. The peak to peak amplitude of this signal or the amplitude of a resonance peak in the cantilever spectrum were used as a measure of the PTIR signal intensity. The resonance spectrum of the cantilever was calculated from the oscillatory decay using the Fast Fourier-Transform (FFT) algorithm.

An OPO (Optical Parametric Oscillator) laser was used as the excitation source. For the measurement of PTIR spectra, the laser was scanned from 900 cm$^{-1}$ to 3000 cm$^{-1}$ in 2 cm$^{-1}$ steps, with a gap in the interval 1800 cm-1 - 2700 cm-1 to reduce measurement time. The plane of polarization was 90 degrees (perpendicular polarization) with a beam incidence angle of about 70 degrees from the normal. Power was set at 17% of the maximum and 512 measurements were co-averaged for each spectral point. The contact resonances were selected using a search location of 184 kHz and a Gaussian filter with a width at half-maximum of 10 kHz. The maximum peak-to-peak amplitude of the oscillatory decay was used for recording PTIR spectra. For the measurement of PTIR maps, the laser was set at the selected wavenumber while the AFM tip was scanned over the sample in contact mode. The contact mode scan was performed using a scan rate of 0.1 Hz to 0.2 Hz. Spatial resolution varied from 256-pixel per line to 512-pixel per line in the X and Y direction using the feedback loop of the Z scanner. The plane of polarization for the incident light was set at 0 or 90 degrees, parallel to the surface. Power was set at 8% to 20% of the maximum and 8 to 16 pulses were co-averaged for each spectral point. The contact resonance was selected using a search location of 64 kHz or 184 kHz via a Gaussian filter with a half-width of 10-50 kHz. Either the maximum peak-to-peak amplitude of the oscillatory decay or the amplitude of the resonance peak in the resonance spectrum were used for PTIR mapping. PTIR maps are not normalized to laser power by the nanoIR2 instrument during collection. AFM images were collected in contact mode using the AFM functionality of the nanoIR2 instrument, either without laser excitation or in conjunction with PTIR imaging. AFM and AFM-IR maps were imported into Gwyddion 2.53 (http://gwyddion.net/) for processing and presentation. All maps were corrected to assign a value of zero to the minimal signal. AFM-IR spectral traces were imported into OriginPro 2019 (Origin Lab) for graphical plotting.

Buccal cells were collected from the cheek epithelium of a volunteer via a buccal swab using a cotton tipped stick and transferred to a ZnSe window and used immediately. Samples were inspected using the optical imaging pathway of the instrument. Only cells with a larger, well defined nucleus were retained. Cells with a micronucleus, double nuclei, or a fragmented nucleus were avoided. Selected cells were used for PTIR experiments without further processing.

## 5. Conclusions

We use PTIR to perform subcellular spectromicroscopy and imaging of intact buccal epithelial cells. The cells are measured in contact mode without being fixed nor actively dried and appear to retain their overall morphology and structure. The experiments demonstrate the possibility to obtain useful IR spectroscopic information on subcellular structures in cells without the use of fixation or extensive drying, thus avoiding the complications of measurements in aqueous solution. We show that the cells retain their structure despite being exposed to the mechanical and thermal perturbations associated to the PTIR measurements. The approach can be easily extended to other cellular systems and to tissue samples, thus providing a useful platform for spectroscopic studies of functional biological samples with spatial resolution unaffected by the diffraction limit.

We show that measurements on intact cells can resolve a multitude of subcellular structures in PTIR imaging and we can collect PTIR spectra of subcellular regions. However, image contrast and spectral band intensity cannot be explained purely in terms of differences in chemical composition and infrared light absorption. Most of the signal in PTIR images comes from spheroidal organelle-like structures, possibly peroxisomes, lysosomes, and mitochondria. Other subcellular structures, such as the cytoskeleton, Golgi and ER membranes, cannot be clearly observed. Similarly, in spectromicroscopy measurements we also report selective response from only some molecular bands, mostly from polypeptides and acyl lipids, which contrasts with the known complexity of cellular IR absorption spectra. PTIR spectral bands from organelles are stronger than those recorded from other regions of the cell, in agreement with the conclusions from imaging experiments. This selective response from specific organelles does not seem to correlate to compositional differences in a simple way. We explore the possibility that the properties of intact cells introduce specific effects in PTIR signal generation, possibly related to the presence of residual water in the sample. It is not clear at present if the peculiar photothermal response of intact cells arises from compositional, thermomechanical, electrodynamic or geometrical factors, or a combination of the same. Answering this question will be the subject of future investigations.

**Supplementary Materials:** The following are available online at www.mdpi.com/xxx/s1.

**Funding:** The research was performed using equipment purchased in the frame of the project co-funded by the Małopolska Regional Operational Programme Measure 5.1 Krakow Metropolitan Area as an important hub of the European Research Area for 2007–2013, project No. MRPO.05.01.00-12-013/15.

This work was in part supported by an OPU16 grant to Luca Quaroni from the National Science Center Poland under contract UMO-2018/31 / B / NZ1 / 01345.

The author is grateful to his wife for financial support during the writing up of the manuscript.

**Acknowledgments:** The author thanks prof. dr. hab. Czesława Paluszkiewicz, prof. dr. hab. Wojciech Kwiatek, dr. Natalia Piergies, of IFJ-PAN for instrumentation access and support, Dr. Miriam Unger, Dr. Anirban Roy, and the staff of Anasys/Bruker for helpful discussion and instrumentation advice.

**Conflicts of Interest:** The author declares no conflicts of interest.

**Appendix A: Comparison of different cells**

Supporting Figure 1 compares PTIR measurements on two different cells using the same excitation wavelength. The cells differ mostly for the size of the nucleus, which is flat and broader in panel A while it is rounded and compact in panel B. The thickness of the nuclei defines the highest position of the cell and allows us to compare signal intensity and sample thickness. It is assumed that the volume of the cell under the tip is dominated by the nucleus when the tip is located on the top of the cell. The ratio of the cell height between A and C is 0.61 and comparable to the ratio of PTIR intensity in the same location, approximately 0.68, showing that a linear relationship between thickness and intensity is retained at least in the nuclear position.

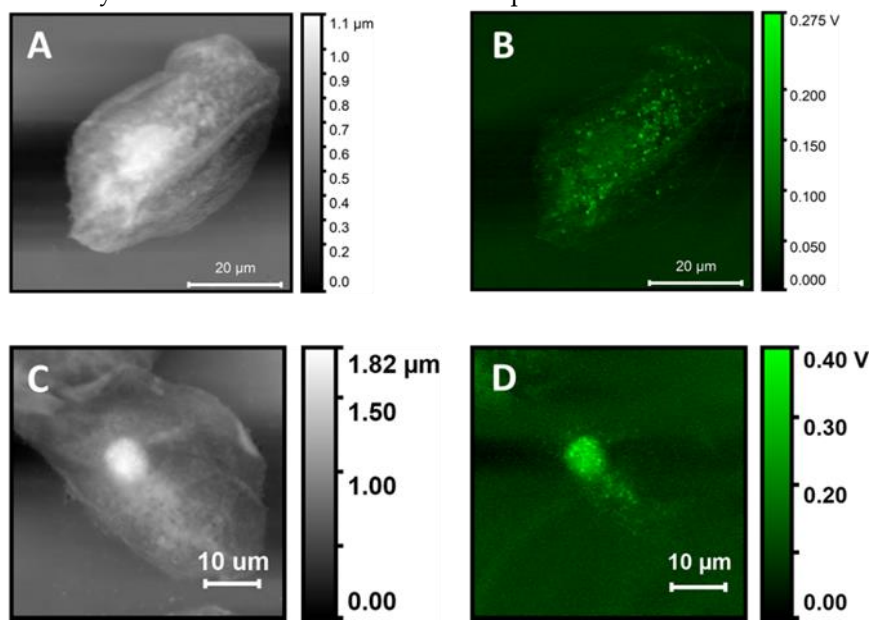

Appendix Figure 1. Comparison of AFM height and PTIR images for two different buccal cells. A,C - Height Maps. B,D PTIR maps at 1650 cm-1.

**Appendix B: Comparison of contact resonance frequency and IR peak map.**

Changes in contact frequency throughout the sample affect the intensity of the PTIR signal, leading to a corresponding modulation of the IR peak and IR amplitude maps. Supporting Figure 2 shows how the changes in a contact frequency map (Panel A) give rise to similar patterns in a PTIR map (Panel B) of a buccal cell. These patterns have a high apparent spatial resolution, comparable to tip size. However, this does not necessarily correspond to the spatial resolution of the spectroscopic information.

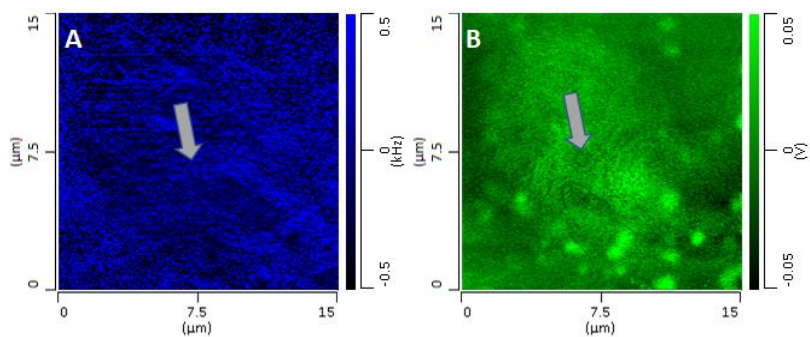

Appendix Figure 2. Comparison of (A) contact frequency map and (B) PTIR intensity map with 1678 cm-1 excitation. The arrows point at locations where a circular structure can be seen in both maps.

**Appendix C: Stability of the Cellular System.**

The stability of the cellular system at ambient conditions was investigated by repeated measurements over a few hours. Appendix Figure 3 compares AFM height and PTIR images collected with 2925 cm$^{-1}$ excitation 8.5 hours apart in the same region of the cell. The overall stability is remarkable and surprising. Height images indicate some minor contraction of the cell, and PTIR images show a minor decrease of signal intensity. Both effects may be due to water loss from the cell.

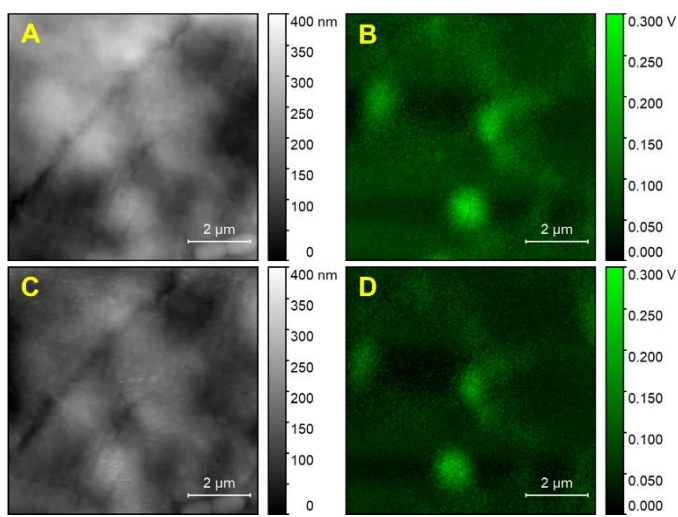

Appendix Figure 3. Comparison of AFM (A, C) and PTIR images (B, D) of a buccal cell kept at approx. 55% humidity for a few hours. A - AFM height at start; B - PTIR with 2925 cm$^{-1}$ excitation at start; C - AFM height after 8.5 h; D - PTIR with 2925 cm$^{-1}$ after 8.5 h.

# References

1. Dürig, U.; Pohl, D.W.; Rohner, F. Near-field optical-scanning microscopy. J. Appl. Phys. 1986, 59, 3318–3327.
2. Cricenti, A.; Generosi, R.; Barchesi, C.; Luce, M.; Rinaldi, M.; Coluzza, C.; Perfetti, P.; Margaritondo, G.; Schaafsma, D.T.; Aggarwal, I.D.; et al. First experimental results with the free electron laser coupled to a scanning near-field optical microscope. Phys. Status Solidi Appl. Res. 1998, 170, 241–247.
3. Knoll, B.; Keilmann, F. Near-field probing of vibrational absorption for chemical microscopy. Nature 1999, 399, 7–10.
4. Dazzi, A.; Prazeres, R.; Glotin, F.; Ortega, J.M. Local infrared microspectroscopy with subwavelength spatial resolution with an atomic force microscope tip used as a photothermal sensor. Opt. Lett. 2005, 30, 2388.
5. Hammiche, A.; Pollock, H.M.; Reading, M.; Claybourn, M.; Turner, P.H.; Jewkes, K. Photothermal FT-IR Spectroscopy: A Step Towards FT-IR Microscopy at a Resolution Better Than the Diffraction Limit. Appl. Spectrosc. 1999, 53, 810–815.
6. Katzenmeyer, A.M.; Holland, G.; Chae, J.; Band, A.; Kjoller, K.; Centrone, A. Mid-infrared spectroscopy beyond the diffraction limit via direct measurement of the photothermal effect. Nanoscale 2015, 7, 17637–17641.
7. Rajapaksa, I.; Uenal, K.; Wickramasinghe, H.K. Image force microscopy of molecular resonance: A microscope principle. Appl. Phys. Lett. 2010, 97, 3–5.
8. Dazzi, A.; Prater, C.B. AFM-IR: Technology and applications in nanoscale infrared spectroscopy and chemical imaging. Chem. Rev. 2017, 117, 5146–5173.
9. Mayet, C.; Dazzi, A.; Prazeres, R.; Ortega, J.-M.; Jaillard, D. In situ identification and imaging of bacterial polymer nanogranules by infrared nanospectroscopy. Analyst 2010, 135, 2540.
10. Dazzi, A.; Prazeres, R.; Glotin, F.; Ortega, J.M.; Al-Sawaftah, M.; de Frutos, M. Chemical mapping of the distribution of viruses into infected bacteria with a photothermal method. Ultramicroscopy 2008, 108, 635–641.
11. Deniset-Besseau, A.; Prater, C.B.; Virolle, M.J.; Dazzi, A. Monitoring TriAcylGlycerols accumulation by atomic force microscopy based infrared spectroscopy in Streptomyces species for biodiesel applications. J. Phys. Chem. Lett. 2014, 5, 654–658.
12. Kochan, K.; Perez-Guaita, D.; Pissang, J.; Jiang, J.H.; Peleg, A.Y.; McNaughton, D.; Heraud, P.; Wood, B.R. In vivo atomic force microscopy-infrared spectroscopy of bacteria. J. R. Soc. Interface 2018, 15.
13. Baldassarre, L.; Giliberti, V.; Rosa, A.; Ortolani, M.; Bonamore, A.; Baiocco, P.; Kjoller, K.; Calvani, P.; Nucara, A. Mapping the amide I absorption in single bacteria and mammalian cells with resonant infrared nanospectroscopy. Nanotechnology 2016, 27, 075101.
14. Kennedy, E.; Al-Majmaie, R.; Al-Rubeai, M.; Zerulla, D.; Rice, J.H. Nanoscale infrared absorption imaging permits non-destructive intracellular photosensitizer localization for subcellular uptake analysis. RSC Adv. 2013, 3, 13789.
15. Kennedy, E.; Al-Majmaie, R.; Al-Rubeai, M.; Zerulla, D.; Rice, J.H. Quantifying nanoscale biochemical heterogeneity in human epithelial cancer cells using combined AFM and PTIR absorption nanoimaging. J. Biophotonics 2015, 8, 133–141.
16. Mayet, C.; Dazzi, A.; Prazeres, R.; Allot, F.; Glotin, F.; Ortega, J.M. Sub-100 nm IR spectromicroscopy of living cells. Opt. Lett. 2008, 33, 1611.
17. Policar, C.; Waern, J.B.; Plamont, M.A.; Clède, S.; Mayet, C.; Prazeres, R.; Ortega, J.M.; Vessières, A.; Dazzi, A. Subcellular IR imaging of a metal-carbonyl moiety using photothermally induced resonance. Angew. Chemie - Int. Ed. 2011, 50, 860–864.
18. Quaroni, L.; Pogoda, K.; Wiltowska-Zuber, J.; Kwiatek, W.M. Mid-infrared spectroscopy and microscopy of subcellular structures in eukaryotic cells with atomic force microscopy-infrared spectroscopy. RSC Adv. 2018, 8, 2786–2794.
19. Perez-Guaita, D.; Kochan, K.; Batty, M.; Doerig, C.; Garcia-Bustos, J.; Espinoza, S.; McNaughton, D.; Heraud, P.; Wood, B.R. Multispectral Atomic Force Microscopy-Infrared Nano-Imaging of Malaria Infected Red Blood Cells. Anal. Chem. 2018, 90, 3140–3148.

20. Ruggeri, F.S.; Marcott, C.; Dinarelli, S.; Longo, G.; Girasole, M.; Dietler, G.; Knowles, T.P.J. Identification of oxidative stress in red blood cells with nanoscale chemical resolution by infrared nanospectroscopy. Int. J. Mol. Sci. 2018, 19, 1–14.
21. Jin, M.; Lu, F.; Belkin, M.A. High-sensitivity infrared vibrational nanospectroscopy in water. Light Sci. Appl. 2017, 6, e17096.
22. Michalczyk, A.; Varigos, G.; Smith, L.; Ackland, M.L. Fresh and cultured buccal cells as a source of mRNA and protein for molecular analysis. Biotechniques 2004, 37, 262–269.
23. Rudney, J.D.; Chen, R. The vital status of human buccal epithelial cells and the bacteria associated with them. Arch. Oral Biol. 2006, 51, 291–298.
24. Lahiri, B.; Holland, G.; Centrone, A. Chemical imaging beyond the diffraction limit: Experimental validation of the PTIR technique. Small 2013, 9, 439–445.
25. Barlow, D.E.; Biffinger, J.C.; Cockrell-Zugell, A.L.; Lo, M.; Kjoller, K.; Cook, D.; Lee, W.K.; Pehrsson, P.E.; Crookes-Goodson, W.J.; Hung, C.S.; et al. The importance of correcting for variable probe-sample interactions in AFM-IR spectroscopy: AFM-IR of dried bacteria on a polyurethane film. Analyst 2016, 141, 4848–4854.
26. Dazzi, A.; Glotin, F.; Carminati, R. Theory of infrared nanospectroscopy by photothermal induced resonance. J. Appl. Phys. 2010, 107.
27. Ruggeri, F.S.; Longo, G.; Faggiano, S.; Lipiec, E.; Pastore, A.; Dietler, G. Infrared nanospectroscopy characterization of oligomeric and fibrillar aggregates during amyloid formation. Nat. Commun. 2015, 6, 7831.
28. Chang, D.C.; Hazlewood, C.F.; Nichols, B.L.; Rorschach, H.E. Spin echo studies on cellular water [15]. Nature 1972, 235, 170–171.
29. Quaroni, L.; Zlateva, T.; Sara, B.; Kreuzer, H.W.; Wehbe, K.; Hegg, E.L.; Cinque, G. Biophysical Chemistry Synchrotron based infrared imaging and spectroscopy via focal plane array on live fi broblasts in D 2 O enriched medium. Biophys. Chem. 2014, 189, 40–48.
30. Daldrop, J.O.; Saita, M.; Heyden, M.; Lorenz-Fonfria, V.A.; Heberle, J.; Netz, R.R. Orientation of non-spherical protonated water clusters revealed by infrared absorption dichroism. Nat. Commun. 2018, 9, 1–7.
31. Khatib, O.; Wood, J.D.; McLeod, A.S.; Goldflam, M.D.; Wagner, M.; Damhorst, G.L.; Koepke, J.C.; Doidge, G.P.; Rangarajan, A.; Bashir, R.; et al. Graphene-Based Platform for Infrared Near-Field Nanospectroscopy of Water and Biological Materials in an Aqueous Environment. ACS Nano 2015, 9, 7968–7975.
32. Meireles, L.; Barcelos, I.; Ferrari, G.A.; Neves, P.A.A. de A.; Freitas, R. de O.; Gribel Lacerda, R. Synchrotron infrared nanospectroscopy on a graphene chip. Lab a Chip - Miniaturisation Chem. Biol. 2019.
33. Vendrell, O.; Gatti, F.; Meyer, H.D. Dynamics and infrared spectroscopy of the protonated water dimer. Angew. Chemie - Int. Ed. 2007, 46, 6918–6921.
34. Kulig, W.; Agmon, N. A "clusters-in-liquid" method for calculating infrared spectra identifies the proton-transfer mode in acidic aqueous solutions. Nat. Chem. 2013, 5, 29–35.
35. Morozovska, A.N.; Eliseev, E.A.; Borodinov, N.; Ovchinnikova, O.S.; Morozovsky, N. V.; Kalinin, S. V. Photothermoelastic contrast in nanoscale infrared spectroscopy. Appl. Phys. Lett. 2018, 112.
36. Nonnenmacher, M.; O'Boyle, M.P.; Wickramasinghe, H.K. Kelvin probe force microscopy. Appl. Phys. Lett. 1991, 58, 2921–2923.
37. Sugimura, H.; Ishida, Y.; Hayashi, K.; Takai, O.; Nakagiri, N. Potential shielding by the surface water layer in Kelvin probe force microscopy. Appl. Phys. Lett. 2002, 80, 1459–1461.
38. Miljković, M.; Bird, B.; Diem, M. Line shape distortion effects in infrared spectroscopy. Analyst 2012, 137, 3954–3964.
39. Laufer, G.; Huneke, J.T.; Royce, B.S.H.; Teng, Y.C. Elimination of dispersion-induced distortion in infrared absorption spectra by use of photoacoustic spectroscopy. Appl. Phys. Lett. 1980, 37, 517–519.